\documentclass[%
 aip,
 amsmath,amssymb,
 reprint,%
]{revtex4-1}

\usepackage{graphicx}% Include figure files
\usepackage{dcolumn}% Align table columns on decimal point
\usepackage{bm}% bold math
\usepackage[utf8]{inputenc}
\usepackage[T1]{fontenc}
\usepackage{mathptmx}
\usepackage{etoolbox}

\usepackage{wasysym}
\usepackage{graphicx} %Include figure files
\usepackage{dcolumn} %Align table columns on decimal point
\usepackage{bm}
\usepackage{color}
\usepackage{txfonts}
\usepackage{microtype}
\usepackage{hyperref}
\usepackage[english]{babel}
\usepackage{slashed}
\usepackage{gensymb}
\usepackage{epsfig}
\usepackage[normalem]{ulem}
\usepackage{amsmath}

\makeatletter
\def\@email#1#2{%
 \endgroup
 \patchcmd{\titleblock@produce}
  {\frontmatter@RRAPformat}
  {\frontmatter@RRAPformat{\produce@RRAP{*#1\href{mailto:#2}{#2}}}\frontmatter@RRAPformat}
  {}{}
}%
\makeatother
\begin{document}

\preprint{AIP/123-QED}

\title{Diagnosis of ultrafast ultraintense laser pulse characteristics by machine-learning-assisted electron spin}

\author{Zhi-Wei Lu}
\thanks{These authors have contributed equally to this work.}
\affiliation{Ministry of Education Key Laboratory for Nonequilibrium Synthesis and Modulation of Condensed Matter, Shaanxi Province Key Laboratory of Quantum Information and Quantum Optoelectronic Devices, School of Physics, Xi'an Jiaotong University, Xi'an 710049, China}
 
\author{Xin-Di Hou}
\thanks{These authors have contributed equally to this work.} 
\affiliation{Ministry of Education Key Laboratory for Nonequilibrium Synthesis and Modulation of Condensed Matter, Shaanxi Province Key Laboratory of Quantum Information and Quantum Optoelectronic Devices, School of Physics, Xi'an Jiaotong University, Xi'an 710049, China}
 
\author{Feng Wan}\email{wanfeng@xjtu.edu.cn}
\affiliation{Ministry of Education Key Laboratory for Nonequilibrium Synthesis and Modulation of Condensed Matter, Shaanxi Province Key Laboratory of Quantum Information and Quantum Optoelectronic Devices, School of Physics, Xi'an Jiaotong University, Xi'an 710049, China}%

\author{Yousef I. Salamin}
\affiliation{Department of Physics, American University of Sharjah, POB 26666, Sharjah, United Arab Emirates}%

\author{Chong Lv}
\affiliation{Department of Nuclear Physics, China Institute of Atomic Energy, P. O. Box 275(7), Beijing 102413, China}%

\author{Bo Zhang}
\affiliation{Key laboratory of plasma physics, Research center of laser fusion, China academy of engineering physics, 621900, Mianshan Rd 64\#, Mianyang, Sichuan, China}%

\author{Fei Wang}
\affiliation{School of Mathematics and Statistics, Xi'an Jiaotong University,
Xi'an, Shaanxi 710049, China}

\author{Zhong-Feng Xu}
\affiliation{Ministry of Education Key Laboratory for Nonequilibrium Synthesis and Modulation of Condensed Matter, Shaanxi Province Key Laboratory of Quantum Information and Quantum Optoelectronic Devices, School of Physics, Xi'an Jiaotong University, Xi'an 710049, China}	

\author{Jian-Xing Li}\email{jianxing@xjtu.edu.cn}
\affiliation{Ministry of Education Key Laboratory for Nonequilibrium Synthesis and Modulation of Condensed Matter, Shaanxi Province Key Laboratory of Quantum Information and Quantum Optoelectronic Devices, School of Physics, Xi'an Jiaotong University, Xi'an 710049, China}	

\date{\today}% It is always \today, today,

\begin{abstract}
Rapid development of ultrafast ultraintense laser technologies continues to create opportunities for studying strong-field physics under extreme conditions. 
However, accurate determination of the spatial and temporal characteristics of a laser pulse is still a great challenge, especially when laser powers higher than hundreds of terawatts are involved. 
In this paper, by utilizing the radiative spin-flip effect, we find that the spin depolarization of an electron beam can be employed to diagnose characteristics of ultrafast ultraintense lasers with peak intensities around $10^{20}$-$10^{22}$~W/cm$^2$.
With three shots, our machine-learning-assisted model can predict, simultaneously, the pulse duration, peak intensity, and focal radius of a focused Gaussian ultrafast ultraintense laser (in principle, the profile can be arbitrary) with relative errors of $0.1\%$-$10\%$.
The underlying physics and an alternative diagnosis method (without the assistance of machine learning) are revealed by the asymptotic approximation of the final spin degree of  polarization.	
Our proposed scheme exhibits robustness and detection accuracy with respect to fluctuations in the electron beam parameters. 
Accurate measurements of the ultrafast ultraintense laser parameters will lead to much higher precision in, for example, laser nuclear physics investigations and laboratory astrophysics studies. Robust machine learning techniques may also find applications in more general strong-field physics scenarios.
\end{abstract}

\maketitle

\section{Introduction}
Recent rapid advances in ultrafast ultraintense laser technology \cite{corde2013femtosecond, mckenna2016high} have opened up broad prospects for vital investigations in laser-plasma physics \cite{esarey2009physics, macchi2013ion, Miao2022Multi}, laser nuclear physics \cite{betti2016inertial, Feng2022Femtosecond}, laboratory astrophysics \cite{remington2006experimental, Lebedev2019Exploring} and particle physics \cite{Di2012Extremely, lu2022Generation}.
In particular, laser systems of peak intensities in the hundreds of terawatts to multi-petawatts have achieved laboratory intensities of the order of $10^{20}-10^{22}$~W/cm$^2$, recently even reaching $\sim 10^{23}$~W/cm$^2$ with a pulse duration of tens-of-femtoseconds \cite{yoon2021realization}. These achievements are paving the way for explorations of strong-field quantum electrodynamics (SF-QED), among other significant applications.
Meanwhile, the unprecedented laser intensities not only cause large fluctuations in the laser output  ($\sim 1\%-20\%$ in the peak power \cite{yoon2021realization}) but also make accurate determination of the laser parameters increasingly difficult.
These parameters play key roles throughout the laser-driven physical processes. 
For instance, in detection of the quantum radiation reaction effects, energy loss of the scattered electron beam serves as the SF-QED signal and is highly correlated with the laser intensity and pulse duration \cite{poder2018experimental,cole2018experimental}.
In the fast ignition of inertial confinement fusion, specific and precise pulse duration and intensity ($\sim10^{20}$ W/cm$^2$) of the ignition laser are required for improving the energy conversion from laser to fuel and suppressing uncertainties in the laser-plasma interactions \cite{betti2016inertial,tabak1994ignition}.
In laser-plasma acceleration, the peak intensity and pulse duration affect the electron and proton acceleration efficiency and stability \cite{fuchs2006laser,bartal2012focusing,simpson2021scaling}. 
Uncertainties in the focal spot, pulse duration, and intensity of the laser pulse can lead to significant deviations from the parameters present in experiments.
Thus, accurate determination of the spatiotemporal properties of the ultrafast ultraintense laser pulses is a fundamental concern for today's laser-matter interaction experiments.

Current schemes to measure the laser spatiotemporal characteristics are based on separate measurements of the focal spot radius (spatial) and the pulse duration (temporal) under low pulse energy, which can minimize the damage to the optical instruments used, followed by extrapolation of the results to the case of full laser power \cite{vais2020characterizing, ciappina2020focal, harvey2018situ}. Due to the nonlinear effects in the amplification and focusing systems, however, the laser intensity obtained with this method may significantly deviate from the exact value \cite{pretzler2000angular, pariente2016space, li2017degradation}.
By comparison, more reliable parameter diagnosis may be achieved via laser-matter interactions, making it possible to directly extract the spatial and temporal information of the ultrafast ultraintense ($I_0 \gtrsim 10^{20}$ W/cm$^{2}$) laser pulses. Three  mainstream diagnostic mechanisms are currently in use.
First, atomic tunneling ionization, in which nonlinear dependence of the multiple-tunneling-ionization rate on the field strength can only be used to diagnose the laser peak intensity with accuracy of $\lesssim30\%\sim50\%$. However, the barrier suppression effect destroys the accuracy and the atom species should be carefully chosen to match the laser intensity requirements \cite{ciappina2019progress, ciappina2020focal, ciappina2020diagnostics}. 
Second, vacuum acceleration of charged particles, in which the laser peak intensity, focal spot size, and pulse duration can be retrieved from the particle spectral analysis. Here, though, the prepulse and plasma effects and the low statistics substantially influence the final spectra and, therefore, one still needs more elaborate considerations \cite{ivanov2018accelerated, vais2018direct, krajewska2019high, vais2020characterizing, vais2020complementary}.
Third, SF-QED effects, e.g., predict the laser intensity and pulse duration separately via analyzing the spectra of electrons \cite{li2018single, mackenroth2019ultra}, photons \cite{har2012peak, harvey2018situ, he2019towards,mackenroth2019determining} and positrons \cite{aleksandrov2021pair}, with detection accuracy of the order of $10\%-50\%$ for laser intensities within the range of $10^{20}-10^{23}$ W/cm$^2$.
Apparently, these methods either require separate diagnoses or can only measure low-precision laser parameter values (the inaccuracy can reach $\simeq 50\%$).
Thus, new detection methods which can achieve high accuracy and simultaneously diagnose the laser intensity, pulse duration, and focal information, are still in great demand.

Recent studies have indicated that spin polarization of the electrons is sensitive to the field strength and profile of the intense laser pulse and, thus, can be manipulated by a laser pulse via the radiative spin-flip effect \cite{li2019ultrarelativistic,song2019spin,li2020polarized}.
These findings have motivated us to explore the possibilities of decoding the pulse information from the spin-polarization of the laser-scattered electron beam.

For decades now, machine learning (ML) techniques have been widely used in particle physics \cite{Karagiorgi2022machine} and astrophysics \cite{vanderplas2012introduction}, with their impact continuously growing on multiscale, highly nonlinear physics such as condensed matter physics and quantum materials science \cite{Carleo2019machine, Schleder2019dft, Carrasquilla2020machine}. ML-assisted methods are more specialized in comprehending multi-modal data (acoustic, visual, and numerical) and optimizing nonlinear extreme physical systems than humans \cite{raghu2020survey} and, thus, can save much time and human effort when integrated into working practices \cite{martin2018automated, hatfield2019blind}.
In particular, the data-driven methods are reshaping our exploration of extreme physical systems, e.g., interaction of the ultrafast ultraintense laser with materials \cite{hatfield2021data}. 
These extreme conditions in the laboratory millimeter-sized plasmas are epitomes of astrophysical scenarios \cite{biener2009diamond}. 
Large quantities of data from such experiments or simulations need to be systematically managed. For instance, around 150 GB of data can be generated in each shot of the National Ignition Facility (NIF) and over 70 GB per minute in the Linac-Coherent-Light-Source (LCLS) \cite{macdonald2016measurement}.
Handling data this size, from both experiments and simulations, is reaching the limits of conventional methods and can obscure the physics behind them. 
By contrast, the ML-assisted method can be data-driven and run in parallel on large-scale CPU or GPU platforms to extract internal correlations between the desired physical quantities.

\begin{figure}[t]
	\includegraphics[width=1\linewidth]{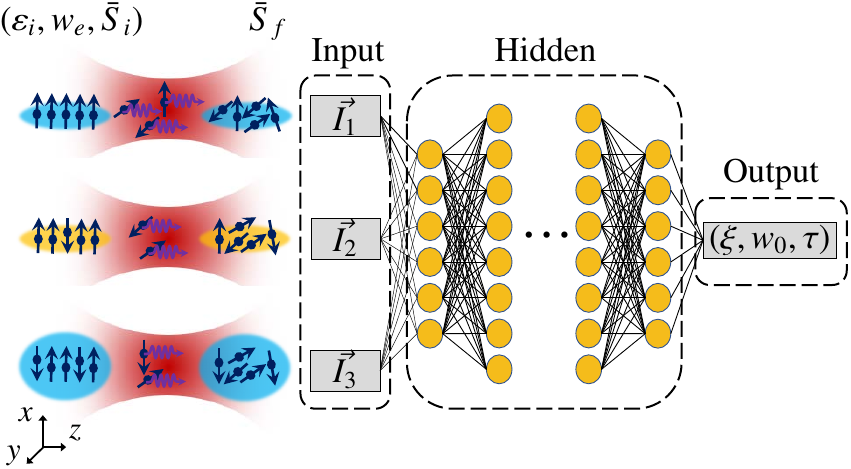}
	\caption{Left: Three different electron beams with parameters ($\varepsilon_i, w_e, \bar{S}_i$) scatter off the same laser pulse and produce final spin degrees of polarization $\bar{S}_{f}$. Right: Topology of the back-propagation neural network (BPNN) used for the parameter prediction which takes the $\vec{I}_{j;j=1,2,3}=[\varepsilon_i, w_e, \bar{S}_i, \bar{S}_f, {\rm ln}(\bar{S}_f/\bar{S}_i)]$ as input data, and produces $(\xi, w_0, \tau)$ as output; see details in Sec.~\ref{twoB}.
	}\label{fig1}
\end{figure}

In this paper, we propose an ML-assisted method to directly diagnose the spatiotemporal characteristics (peak intensity, focal spot size, and pulse duration) of a linearly polarized (LP) laser pulse, based on the spin-analysis of nonlinear Compton-scattered electron beams. 
The interaction scenario and framework of the ML-assisted diagnosis method are shown in Fig.~\ref{fig1}. 
When a transversely polarized (probe) beam of electrons (mean energy $\varepsilon_i$, beam radius $w_e$, and degree of polarization $\bar{S}_i$) propagates along the $z$ direction and collides with the ultrafast ultraintense laser pulse to be diagnosed, electrons can undergo strong nonlinear Compton scattering (NCS) \cite{Di2012Extremely}. 
Due to the radiative spin-flip effect \cite{li2019ultrarelativistic,seipt2019ultrafast,song2019spin}, the degree of polarization  changes from an initial $\bar{S}_i$ to a final $\bar{S}_f$. The differences (i.e., degree of depolarization) $\delta \bar{S} \equiv \bar{S}_i - \bar{S}_f$ from three different beams may be used to determine the laser pulse parameters: normalized intensity $\xi \equiv eE_0/m\omega_0$, focal radius $w_0$, and pulse duration $\tau$, where $-e$ and $m$ are the charge and mass of the electron, $E_0$ and $\omega_0$ are the electric field strength and frequency of the laser field, respectively.
Relativistic units with $c=\hslash=1$ will be used throughout.
In addition to those fixed laser parameters, $\delta \bar{S}$ is related to the spatial distribution (beam radius $w_e$), average energy $\varepsilon_i$, and initial degree of polarization $\bar{S}_i$ of the electron beam.   
However, a one-to-one mapping between the beam parameters $(\varepsilon_i, w_e, \bar{S}_i, \bar{S}_f)$ and the laser parameters $(\xi, w_0, \tau)$ can be a formidable task, because only one output is of relevance, i.e., $\bar{S}_f$.
In order to determine the three unknown laser parameters $(\xi, w_0, \tau)$ simultaneously, at least three sets of output values of $\bar{S}_f$ are required.
Therefore, three independent beams with different parameter combinations are employed here.
These complex multidimensional relationships can be properly handled by the Neural Network  topology shown in Fig.~\ref{fig1}.
Note that this method can induce a spin depolarization of $\simeq 30\%$ for 1-GeV electrons, and $\simeq 40\%$ for 2-GeV ones (laser parameters $\xi \simeq 80$ and $\tau = 14T_0$).
Currently, available spin polarimetries for electrons are based on Mott scattering \cite{Mott1929}, M$\text{\o}$ller scattering \cite{Cooper1975}, linear Compton scattering \cite{Klein1929}, or more efficient NCS \cite{Li2019}. Some recent studies indicate that the detection precision of NCS-based polarimetry can reach about 0.3\% \cite{Li2019}, which qualifies the spin-based method as a new type of high-accuracy diagnostic scheme for ultrafast ultraintense laser pulses.

In Sec.~\ref{two}, a brief description of the Monte-Carlo (MC) simulation method of spin-resolved  NCS will be given, together with the simulation parameters. This is followed by introducing our laser-parameter retrieval technique based on the ML algorithms (see Fig.~\ref{fig1}) and the associated asymptotic formulas. Numerical results and a brief discussion will be given in Sec.~\ref{three}. Our conclusions will be presented in Sec.~\ref{four}.

\section{Spin-based Laser-parameter diagnostic methods}\label{two}

As an illustrative example, diagnosis of a tightly focused laser with a double-Gaussian (spatial and temporal) distribution is considered.
In principle, the envelope of the laser can be arbitrary, but should be predetermined via experimental methods, for instance, from a low-power splitting beam.
Once the envelope form is known, the following methods can be used to retrieve the laser pulse parameters from the spin diagnosis of the scattered electrons.

\subsection{Spin-resolved NCS and interaction scenario}\label{twoA}
Our analysis of the radiative spin-flip effect is based on MC simulation method proposed in~\cite{li2019ultrarelativistic, xue2020generation}, in which the spin-resolved probability of NCS in the laser-beam scattering is considered in  the local constant field approximation (LCFA) \cite{katkov1998electromagnetic, li2019ultrarelativistic}. 
After emission of a photon, the electron spin state collapses into one of its basis states defined with respect to an instantaneous spin quantization axis (SQA) chosen along the magnetic field in the rest frame of the electron.
In Fig.~\ref{fig1}, the laser is linearly polarized along the $x$-direction, so its magnetic field component is $B_y$. The SQA tends to be anti-parallel to the magnetic field in the rest frame of the electron. Depolarization amounts to the electron spin acquiring a certain spin polarization in the $y$-direction, which gets cancelled from the net polarization by the periodic magnetic field, i.e.,  $\bar{S}_{f,y}\approx0$. 
Therefore, we focus our analysis, in what follows, on the electron polarization in the $x$-direction.
In NCS, the invariant parameter characterizing the quantum effects is $\chi\equiv e\sqrt{-(F_{\mu\nu}p^{\nu})^2}/m^3$~\cite{ritus1985quantum, katkov1998electromagnetic}, where $F_{\mu\nu}$ and $p^{\nu}$ denote the electromagnetic field tensor and the four-momentum of the electron, respectively. 
In a colliding geometry, $\chi\approx 2\xi\gamma_e\omega_0/m$, where $\gamma_e$ denotes the electron's Lorentz factor.
To excite the radiative spin-flip process, $\chi$ should be in the range of 0.01 to 1, over which the nonlinear Breit-Wheeler pair-production can be suppressed.

The LP laser parameter set for the training data includes: wavelength $\lambda_0=0.8$ $\mu {\rm m}$, focal radius $w_0=[2,3,4,5]\lambda_0$, peak intensity $\xi =[10,15,20,30,40,45,60,80]$, and pulse duration $\tau=[2,6,10,14]T_0$, with $T_0$ denoting the laser period.
The probe electron beam has a polar angle $\theta_e=\pi$, azimuthal angle $\phi_e=0$, and angular divergence $\sigma_\theta = 0.3~\mathrm{mrad}$. The initial kinetic energies are $\varepsilon_i=[0.5,1,1.5,2]$ GeV, with relative energy spread $\sigma_\varepsilon/\varepsilon_i=0.05$, and initial average degree of spin-polarization along the $x$-direction $\bar{S}_{i,x}=[0.6, 0.8, 1.0]$  (here, $\chi_{\rm max}\lesssim1$, i.e., the pair-production effect on the final electron distribution is negligible for the present parameters). The beam radius $w_e=[1,2,3,4]\lambda_0$, beam length $L_e = 5 \lambda_0$, and the total number of electrons is $5\times10^5$ with transversely Gaussian and longitudinally uniform distributions, attainable by current laser wakefield accelerators \cite{esarey2009physics}. 

\subsection{Neural Network assisted diagnosis}\label{twoB}

Decoding the spatiotemporal characteristics of the ultrafast ultraintense laser from information carried by the scattered electron beam is an inverse transformation that requires multidimensional input and output. To make full use of the electron beam data, we build a standard BPNN via $\texttt{PyTorch}$ to train and predict the scattering laser parameters \cite{Paszke2019pytorch}.
The input data is composed of the energy, beam radius, initial and final average spins, and logarithm of the ratio of final spin to initial spin of the electron beam, in the vector $\vec{I} \equiv [\varepsilon_i, w_e, \bar{S}_i, \bar{S}_f, \ln (\bar{S}_f/\bar{S}_i)]$; see Fig.~\ref{fig1}.
About 1000 sets of input data are obtained via the MC simulation and rearranged/recombined to about $3\times10^4$ sets for training.
Then the input data $(\vec{I_1}, \vec{I_2}, \vec{I_3})$ is normalized via $\texttt{StandScaler}$ function. After random permutation, the input information is preprocessed by the second-order polynomial feature function ($\texttt{PolynomialFeatures}$) to construct implicit connections between them.

\begin{figure}[t] 
	\begin{center}
		\includegraphics[width=1\linewidth]{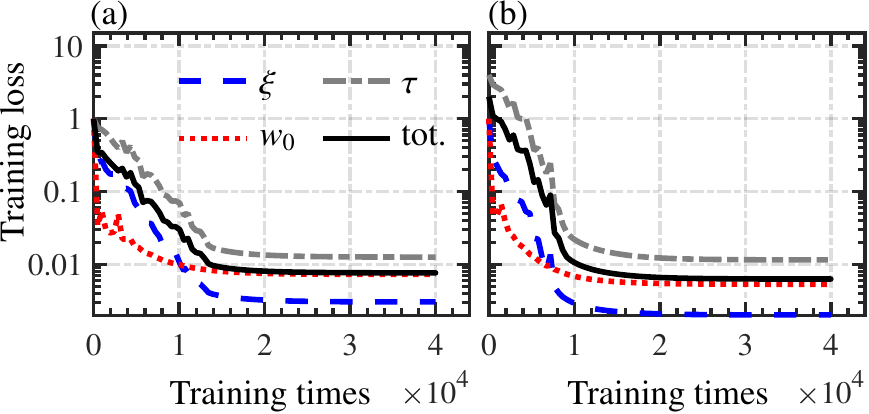}
		\caption{(a) and (b): Training loss (mean squared errors for all training samples) evolutions of $\xi$, $w_0$, $\tau$ and the total loss (tot.) vs training times. Learning ratios of $\xi, w_0$ and $\tau$ are 1:1:1 in (a), and 1:1:2 in (b).}\label{fig2}
	\end{center}
\end{figure}

In our BPNN, we choose eight fully connected hidden layers and the corresponding numbers of nodes are (128, 256, 512, 512, 512, 512, 256, 128). The numbers of hidden layers and nodes here ensure adequate prediction accuracy and appropriate computing resources. 
The activation functions alternatively use $\texttt{tanh}$ and $\texttt{PReLU}$ between different layers.
Mean squared error ($\texttt{MSELoss}$) is used as the loss function, and the stochastic gradient descent ($\texttt{SGD}$) method is used as the optimizer.
After each training iteration, the optimizer clears old gradients, and losses are back-propagated for the calculation of new gradients. Finally, the network parameters are updated according to the new gradients.
The initial learning rate is set as 0.3 and the adjustment factor of the exponential learning rate ($\texttt{ExponentialLR}$) scheduler is set as 0.9.
In our calculations, the total number of training iterations is $4\times10^4$.  
In order to enhance the learning efficiency of the model on the laser pulse duration $\tau$, we set the learning ratios of  $\xi$, $w_0$ and $\tau$ as 1:1:1 and 1:1:2; see Figs.~\ref{fig2} (a) and (b) in the two separate models, respectively. 
Note that the training loss measures the training efficiency of the model. The training loss may increase due to the inappropriate network structure design and will decrease due to effective learning. In the final stable stage, there may be over-fitting to the training data. However, the over-fitting can be restrained by using a technique such as weight upper limit \cite{srebro2005rank} or dropout \cite{srivastava2014dropout}.
For instance, the losses of $\xi$, $w_0$, and $\tau$ are reduced for the learning ratios of 1:1:2, and further increasing the ratio of $\tau$ will produce larger losses in other parameters. 
This BPNN model will be used in the latter prediction. 
In principle, the ML-assisted method is not limited to the current application, but can also be used for other inverse problems.

\subsection{Analytical asymptotic models}\label{twoC}

Asymptotic estimation of the depolarization effect is done below analytically from the radiative equations of motion for the dynamics [Landau-Lifshitz (LL) equation \cite{landau1975lifshitz}] and the spin [modified Thomas-Bargmann-Michel-Telegdi (T-BMT) equation \cite{guo2020stochasticity}]. Dependence of the spin dynamics on the electron energy follows assuming weak radiation. Then the quantum-corrected LL equation is used to obtain the approximated electron energy, which is then plugged into the solution for spin dynamics.

The radiative spin evolution is composed of the Thomas precession (subscript ``T'') and radiative correction terms (subscript ``R'') \cite{guo2020stochasticity}. That evolution is governed by 

\begin{subequations}\label{spin1}
	\begin{eqnarray}
	\frac{{\rm d}{\bf S}}{{\rm d}\eta}&=&\left(\frac{{\rm d}{\bf S}}{{\rm d}\eta}\right)_{T}+\left(\frac{{\rm d}{\bf S}}{{\rm d}\eta}\right)_{R} ,\\
	\left(\frac{{\rm d}{\bf S}}{{\rm d}\eta}\right)_{T}&=&\frac{e\gamma_e}{\left(k\cdot p_i\right)}{\bf S}\times\left[-\left(\frac{g}{2}-1\right)\frac{\gamma_e}{\gamma_e+1}\left({\bm \beta}\cdot{\bf B}\right)\cdot{\bm \beta}\right.\nonumber \\
	&&\left.+\left(\frac{g}{2}-1+\frac{1}{\gamma_e}\right){\bf B}-\left(\frac{g}{2}-\frac{\gamma_e}{\gamma_e+1}\right)\bm{\beta}\times{\bf E}\right], \\
	\left(\frac{{\rm d}{\bf S}}{{\rm d}\eta}\right)_{R}&=&-P\left[\psi_1(\chi){\bf S}+\psi_2(\chi)({\bf S}\cdot{\bm \beta}){\bm \beta}+\psi_3(\chi){\bf n}_B\right],
	\end{eqnarray}
\end{subequations}
where ${\bf E}$ and  ${\bf B}$ are the  laser electric and magnetic fields, respectively, $p_i$, $k$, $\eta$ and $g$ are: the electron momentum 4-vector, the laser wavevector, the laser phase, and the electron gyromagnetic ratio, respectively, $P=\alpha_f m^2/[\sqrt{3}\pi\left(k\cdot p_i\right)]$, $\psi_1(\chi)$ = $\int_{0}^{\infty} u''{\rm d}u {\rm K}_{\frac{2}{3}}(u')$, $\psi_2(\chi)$ = $\int_{0}^{\infty} u''{\rm d}u \int_{u'}^{\infty}{\rm d}x{\rm K}_{\frac{1}{3}}(x)$-$\psi_1(\chi)$, $\psi_3(\chi)$ = $\int_{0}^{\infty} u''{\rm d}u {\rm K}_{\frac{1}{3}}(u')$,  $u'=2u/3\chi$, $u''=u^2/(1+u)^3$, $u=\varepsilon_\gamma/\left(\varepsilon_0-\varepsilon_\gamma\right)$,  $\varepsilon_0$ and $\varepsilon_\gamma$ are the electron energy before radiation and the emitted photon energy, respectively, ${\rm K}_n$ is the $n$th-order modified Bessel function of the second kind, and $\alpha_f=1/137$ is the fine structure constant. The SQA is chosen along the magnetic field ${\bm n}_B={\bm\beta}\times\hat{\bm{ a}}$, with ${\bm \beta}=\bm{v}/c$ the scaled electron  velocity and $\hat{\bm{ a}}=\bm{a}/|\bm{a}|$ a unit vector along the electron acceleration  $\bm{ a}$.

To facilitate theoretical analysis and extract analytical formulas, some approximations will be made with current laser and electron beam parameters in mind, i.e., GeV electron beam interacting with an LP laser ($\xi<100$) and $0.1 \lesssim \chi \lesssim 1$. 
Due to laser defocusing, the Thomas-term-induced variation $\delta S_T$ is $\lesssim 10^{-4}$, and only the dominant term, i.e., the radiative correction will be considered.
Furthermore, initial velocity of the electron beam is along the $z$ direction, with $\beta_z\gg\beta_x(\beta_y)$, thus the $\psi_2$-term is negligible for initial TSP electrons. Moreover, due to the periodic nature of the magnetic field, contribution of the $\psi_3$-term vanishes on average within one laser period. 
Hence, approximate evolution of the spin components may be obtained from
\begin{subequations}\label{spin2}
	\begin{eqnarray}
	\frac{{\rm d}{S_x}}{{\rm d}\eta}&\simeq&\frac{C\psi_1(\chi)}{\gamma_e}S_x, \label{spin2a} \\
	\frac{{\rm d}{S_y}}{{\rm d}\eta}&\simeq&\frac{C\psi_1(\chi)}{\gamma_e}S_y, \label{spin2b} \\
	\frac{{\rm d}{S_z}}{{\rm d}\eta}&\simeq&\frac{C(\psi_1(\chi)+\psi_2(\chi))}{\gamma_e}S_z, \label{spin2c}
	\end{eqnarray}
\end{subequations}
where $C=- \frac{\alpha_f}{2\sqrt{3}\pi} \frac{\omega_0}{m}$. 
Because $\psi_{1}(\chi)>0$ and $\psi_{2}(\chi)<0$, depolarization in the $x$- and $y$-directions is faster than in the $z$-direction. For instance, for a laser with parameters of $\xi=60$, $\tau=8T_0$, and $w_0=5\lambda_0$, and the electron beam of  Fig.~\ref{fig4} (a), the final average spin degrees of polarization are $S_{x,f}\approx0.8201$, $S_{y,f}\approx0.8211$ and $S_{z,f}\approx0.8741$, for $\bar{S}_{i,x}=1$, $\bar{S}_{i,y}=1$, and $\bar{S}_{i,z}=1$. Thus, in this paper, we take the electron beam initially polarized along the $x$-direction for a larger detection signal.

Under the assumption of weak radiation loss $\frac{{\rm d}{{\gamma}_e}}{{\rm d}\eta} \simeq 0$ and $\chi(\eta) \simeq 2 \frac{\omega_0}{m}\gamma_e\xi \sin^2(\eta)$, one can obtain, to leading-order approximation, $\psi_1(\chi) \simeq f_1\chi^2$, for $0.1\lesssim \chi \lesssim 1$ and $f_1 \approx 0.25$, is obtained via curve fitting. Integrating Eq.~(\ref{spin2a}), the asymptotic $\bar{S}_{f,x}$, using the laser-beam parameters, will be given by
\begin{eqnarray}\label{solution0}
\ln \frac{\bar{S}_{f,x}(\tau)}{\bar{S}_{i,x}(0)}&\simeq& M_1\gamma_e\xi^2\tau,
\end{eqnarray}
with the factor $M_1= -\frac{\sqrt{3}}{2} \alpha_f \frac{\omega_0}{m} f_1 \approx -4.81\times10^{-9}$ and $\tau$ is the pulse duration in units of the laser period $T_0$.

To be precise, the radiated photon energy (radiation loss $\overline{\varepsilon}_{\gamma}$) should be taken into account for $0.1 \lesssim \chi \lesssim 1$. 
Here, we use the quantum-corrected LL equation to include the radiation loss \cite{niel2018quantum} via
\begin{subequations}\label{force}
	\begin{eqnarray}
	\frac{{\rm d}{\bf P}}{{\rm d}t}&=&{\bf F}_L+{\bf F}_{\rm rad}, \\
	{\bf F}_{\rm rad}&=&-C'\chi^2\mathcal{G}(\chi){\bm \beta}/({{\bm \beta}}^2),
	\end{eqnarray}
\end{subequations}
where ${\bf F}_L \equiv q({\bf E} + \bm{ v} \times {\bf B})$ denotes the Lorentz force and ${\bf F}_\mathrm{rad}$ the radiation reaction force, $C'=2\alpha_f^2m/(3r_e)$, $r_e$ the classical electron radius and $\mathcal{G}(\chi) \simeq [1+4.8(1+\chi)\ln(1+1.7\chi)+2.44\chi^2]^{-2/3}$ the quantum correction function \cite{tamburini2010radiation}. 
For $0.1 \lesssim \chi \lesssim 1$, assuming $\chi(\eta)\simeq 2 \frac{\omega_0}{m} \gamma_e\xi \sin^2(\eta)$ and making the approximation $\chi^2\mathcal{G}(\chi)\simeq f_2\chi^2$ (with a fitting factor of $f_2 \approx 0.077$), the radiation loss (averaged over all electrons, i.e., ignoring the stochastic effect) is given by $\overline{\varepsilon}_{\gamma}=\int_{0}^{\eta}{\rm d}\eta{\bf F}_{\rm rad}\frac{{\rm d}t}{{\rm d}\eta}\simeq M_2\tau\gamma_e^2\xi^2$, where $M_2=\pi \alpha_f \frac{\omega_0}{m} f_2 \approx 5.36\times10^{-9}$. Then, replacing $\gamma_e$ in Eq.~(\ref{solution0}) with $\gamma_e-\overline{\varepsilon}_{\gamma}$, analytical asymptotic estimation of the final spin $\bar{S}_{f,x}$ will be given by
\begin{equation}\label{solution1}
\ln \frac{\bar{S}_{f,x}(\tau)}{\bar{S}_{i,x}(0)} \simeq M_1\gamma_e\xi^2\tau(1-M_2\gamma_e\xi^2\tau).
\end{equation}

\section{Results and discussions}\label{three}

\begin{table}[t]
	\caption{Operational parameters of some international ultrafast ultraintense laser facilities: total energy $E_L$, central wavelength $\lambda$, peak intensity $I_0$, pulse duration $\tau$, and focal radius $ w_0$.}
	\setlength{\tabcolsep}{4pt}
	\label{Tab:facilities}
	\renewcommand{\arraystretch}{1.5}
	\resizebox{0.5\textwidth}{!}{
		\begin{tabular}{c|c|c|c|c|c}
			\hline\hline
			Project & $E_L$ (J) & $\lambda$ ($\mu$m)  &$I_0$ (W/cm$^2$); $\xi$&$\tau$ (fs); ($T_0$)&$w_0~(\lambda)$\\
			\hline
			ELI-NP~\cite{Tanaka2020current} & 20 & 0.82 & $5.6\times10^{21}$; 52.43 & 18.75; 6.86 & 3.63\\
			\hline
			J-KAREN~\cite{Aoyama20030} & 28.4 & 0.8 & $3.8\times10^{21}$; 42.14 & 32.9; 12.33 & 4.75\\
			\hline
			GIST~\cite{Yu2012generation} & 44.5 & 0.81 & $10^{22}$; 69.21 & 30; 11.1 & 3.79\\
			\hline
			SILEX-$\rm\uppercase\expandafter{\romannumeral2}$~\cite{Hong2021commissioning} & 30 & 0.8 & $5\times10^{20}$; 15.28 & 30; 11.24 & 6.16\\
			\hline
			APOLLON~\cite{Burdonov2021characterization} & 10 & 0.815 & $2\times10^{21}$; 31.14 & 24; 8.83 & 2.92\\
			\hline\hline
	\end{tabular}}
\end{table}

\begin{figure}[t] 
	\begin{center}
		\includegraphics[width=0.9\linewidth]{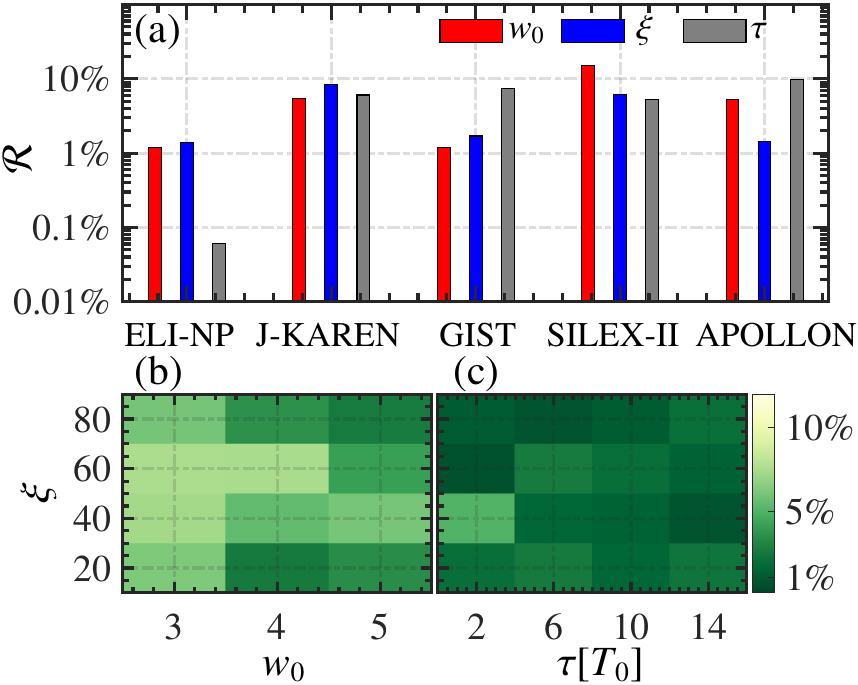}
		\caption{(a) Relative errors $\mathcal{R}=(\mathcal{R}_\xi, \mathcal{R}_\tau, \mathcal{R}_w$) between the predicted and theoretical values of $(\xi,\;\tau,\; w_0)$ for the facilities in Table~\ref{Tab:facilities}, where $\vec{I_1}$, $\vec{I_2}$, and $\vec{I_3}$, respectively, are: $(\varepsilon_i=1$ GeV, $w_e=\lambda_0$, $\bar{S}_{i,x}=1)$, $(\varepsilon_i=1$ GeV, $w_e=3\lambda_0$, $\bar{S}_{i,x}=1)$ and $(\varepsilon_i=1.5$ GeV, $w_e=\lambda_0$, $\bar{S}_{i,x}=1)$. (b) Distribution of the total relative error $\mathcal{R}_1=\sqrt{\mathcal{R}_\xi^2+\mathcal{R}_w^2}$ in the $\xi$-$w_0$ plane, where $\tau=10T_0$ and ($\vec{I_1},~\vec{I_2},~\vec{I_3}$) are the same as in (a). (c) Distribution of the total relative error $\mathcal{R}_2=\sqrt{\mathcal{R}_\xi^2+\mathcal{R}_\tau^2}$ in the $\xi$-$\tau$ plane, where $w_0=5\lambda_0$ and $\vec{I_1}$, $\vec{I_2}$, and $\vec{I_3}$, respectively, are: $(\varepsilon_i=500$ MeV, $w_e=\lambda_0$, $\bar{S}_{i,x}=1)$, $(\varepsilon_i=500$ MeV, $w_e=4\lambda_0$, $\bar{S}_{i,x}=0.8)$, and $(\varepsilon_i=2$ GeV, $w_e=\lambda_0$, $\bar{S}_{i,x}=0.6)$.}\label{fig3}
	\end{center}
\end{figure}
To demonstrate efficiency of the proposed diagnosis method, some operational parameters of petawatt-scale lasers at a number of international facilities will be used; see Table.~\ref{Tab:facilities}. 
The corresponding depolarization processes, investigated via MC simulations, indicate that the relative errors between the predicted and input parameters are of orders $0.1\%$ to $10\%$; see Fig.~\ref{fig3} (a).
After consecutive training, the BPNN model grasps the pattern of the radiative spin-flip effect and, therefore, is capable of accurately predicting the laser characteristics, i.e., $(\xi,\;\tau,\; w_0)$, simultaneously.
Due to limited training data and cycle, the relative prediction errors for $\xi$, $\tau$, and $w_0$ (simultaneously) are of the order of $\mathcal{R}_{1(2)}\lesssim 10\%$; see Figs.~\ref{fig3} (b) and (c).
Compared with cases of $w_0 \gtrsim 3\lambda_0$, the number of electrons scattered by a tightly focused laser ($w_0 \lesssim 3\lambda_0$) is lower due to the small Rayleigh range ($z_R = \pi w_0^2/\lambda$). Thus, the beam-averaged spin-flip effect is relatively more sensitive to variations in the electron beam parameters and the relative error $\mathcal{R}_1$ is larger for $w_0 \lesssim 3\lambda_0$. For a laser radius $w_0\gtrsim5\lambda_0$, already beyond the current training range, certain overfitting is expected. For the SILEX-II, for example, the relative error $\mathcal{R}_w\sim 15\%$. 
By comparison, the prediction error for the laser pulse duration is $\mathcal{R}_2 \lesssim 5\%$, while for most regions, $\mathcal{R}_2 \sim 1\%$, i.e., the prediction of $\tau$ is more accurate than $w_0$; see Figs.~\ref{fig3}(a) and (c).
This scheme is quite stable with respect to fluctuations in the electron beam parameters, as is shown in Fig.~\ref{fig5}.

\begin{figure}[t]
	\begin{center}
		\includegraphics[width=1\linewidth]{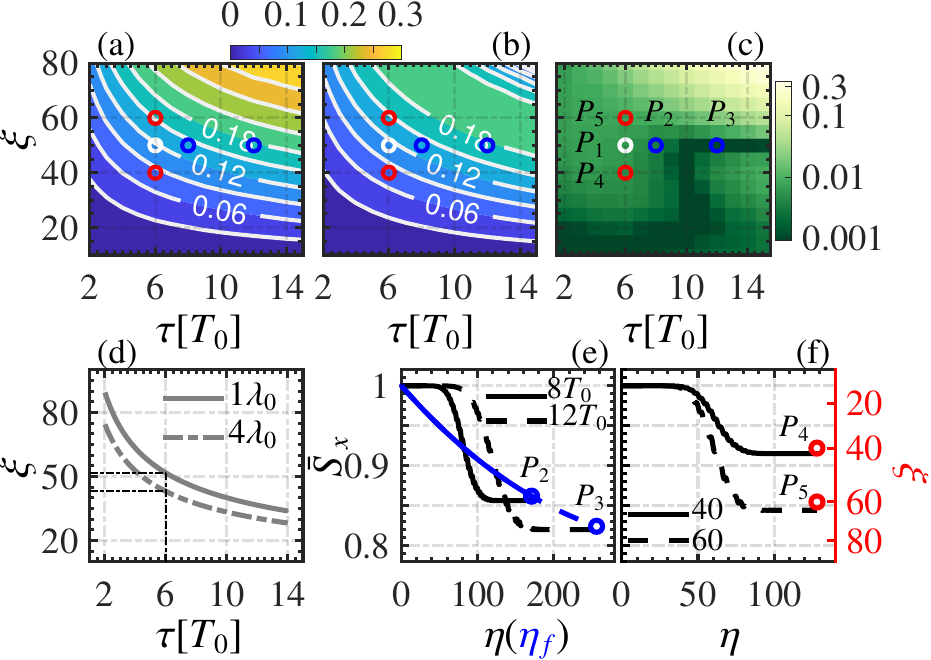}
		\caption{(a) and (b) Transverse spin degree of depolarization of the probe electron beams $\delta{\bar{S}}_x \equiv \bar{S}_{i,x} - \bar{S}_{f,x}$ vs the laser peak intensity $\xi$ and pulse duration $\tau$. For (a) $\delta \bar{S}^{MC}_x$ is calculated by MC method and in (b) $\delta \bar{S}^{AE}_x$ is calculated by asymptotic estimation from Eq.~(\ref{solution1}). Here, the laser radius $w_0=5 \lambda_0$, the probe electron beam energy is $\varepsilon_i=1$ GeV, beam radius $w_e=\lambda_0$, and initial average spin $\bar{S}_{i,x}=1$ are used. Other parameters are the same as in Fig.~\ref{fig3}. (c) Relative error $\mathcal{R}_s=|\delta{\bar{S}}_x^{MC} - \delta{\bar{S}}_x^{AE}|/\delta{\bar{S}}_x^{MC}$ vs $\xi$ and $\tau$. (d) $\delta S(\xi, \tau) = 0.12$ (white circles $P_1(\xi = 50, \tau = 6T_0)$ in (a)-(c)) for $w_e=1\lambda_0$  (solid line) and $w_e=4\lambda_0$ (dash-dotted line). (e) $\bar{S}_x$ vs the laser phase $\eta \equiv \omega_0 (t - z)$. The solid and dashed black lines (averaged MC evolution process) correspond to the blue circles $P_2(\xi = 50, \tau = 8T_0)$ and $P_3(\xi = 50, \tau = 12T_0)$ in (a)-(c), respectively. The blue lines and circles indicate the analytical calculations (only relate to the final laser phase $\eta_f$). (f) $\bar{S}_x$ vs the laser phase $\eta$, The solid and dashed lines correspond to the red circles $P_4(\xi = 40, \tau = 6T_0)$ and $P_5(\xi = 60, \tau = 6T_0)$ in (a)-(c), respectively. Black lines are from the averaged MC evolution calculation and red circles (right axis) are from the analytical calculations.}\label{fig4}
	\end{center}
\end{figure}

Physical essence of the ML-assisted pulse information decoding method can be revealed by our analytical asymptotic estimation on the basis of Eq.~(\ref{solution1}) which is in good agreement with the numerical MC results over a wide range of laser parameter values; see Figs.~\ref{fig4} (a)-(c). The distributions of $\delta{\bar{S}}^{MC,AE}_x$ with respect to $\xi$ and $\tau$ are shown in Figs.~\ref{fig4} (a) and (b), where superscripts ``MC" and ``AE" denote the results from MC and analytical asymptotic estimation (AE) methods, respectively. As expected, $\delta{\bar{S}}_x$ increases as $\xi$ and $\tau$ both increase, and a specific spin change $\delta{\bar{S}}_x$ determines a curve that binds $\xi$ with $\tau$ (or a hyperplane for $\xi$, $\tau$ and $w_0$), i.e., the NCS acts as a nonlinear function $\mathcal{F}(\cdot, \cdot)$ which maps the laser pulse parameters $(\xi, \tau)$ to a degree of depolarization of the electron beam $\mathcal{F}(\xi, \tau) \rightarrow \delta \bar{S}_x$. Quite remarkably, the corresponding relative error $\mathcal{R}_s$ in the parameter ranges of $\xi\in(10,60)$ or $\tau\in(2,6)T_0$ is $\mathcal{R}_s \simeq 1\%$; see Fig.~\ref{fig4}(c). 
With the analytical AE extracted subject to the condition $0.1<\chi\lesssim1$, and for $\xi>60$ and $\tau>6$, the low-order estimation deviates from the MC result, due to the nonlinear radiative effects. 
However, the ML-assisted method is data-driven, i.e., the algorithms can still grasp the correlations between laser pulse parameters and depolarization of the electron beam, without artificial restrictions; see the prediction accuracy (the total relative error $\mathcal{R}_2\sim1\%$) for high laser intensity and long pulse duration in Fig.~\ref{fig3}(c). 

Figure \ref{fig4}(d) illustrates how to determine $\xi$ and $w_0$ via AE for a specific set of parameters ($\xi=50$, $\tau=6T_0$ and $w_0=5\lambda_0$) marked as white circles $P_1$ in Figs.~\ref{fig4} (a)-(c).
Here, the pulse duration $\tau$ is pre-acquired with other diagnostics, for instance, from the low-power mode of detection. This is a restriction not encountered in the ML-assisted method.
Then, a sub-micrometer probe is used to collide with the laser pulse from which one obtains $\delta \bar{S}_1$; see solid line labelled by ``$1\lambda_0$''  in Fig.~\ref{fig4}(d) which has been obtained from Eq.~(\ref{solution1}). 
After that, a second probe with beam radius $w_e = 4\lambda_0$ produces $\delta \bar{S}_2$, the dot-dashed line labelled by ``$4\lambda_0$''  in Fig.~\ref{fig4}(d). 
According to Eq.~(\ref{solution1}), two average intensities $\bar{\xi}_1$ and $\bar{\xi}_2$ can be determined from $\delta \bar{S}_1$ and $\delta \bar{S}_2$, corresponding to different beam radii, respectively.
Since $w_0\gg w_e$, the average laser intensity sensed by the sub-micrometer probe can approximately serve as the peak intensity in the focusing region. Thus, $\bar{\xi}_1=51.62$ is identified as the peak intensity of the laser pulse, with a relative error of $3.2\%$. Whereas,
$\bar{\xi}_2=42.96$, corresponding to $w_e = 4\lambda_0$, is taken as the average intensity within the probe radius, i.e.,  $\bar{\xi}_2=\bar{\xi}_1\int_{-w_e}^{w_e}e^{-r^2/w_0^2}dr$.
Numerical calculation gives the focal radius $w_0=5.18 \lambda_0$, with a relative error of $3.6\%$. 
Note that, in Eq.~(\ref{solution1}), once $\tau$ (or $\xi$) is given, the map between $\delta \bar{S}$ and the other parameter is uniquely fixed.
For instance, in Fig.~\ref{fig4} (d), once $\xi$ is fixed (points $P_2$ and $P_3$ in Figs.~\ref{fig4} (a)-(c)), there will be only one intersection (the final phase $\eta_f$) between Eq.~(\ref{solution1}) and the temporal evolution of the average spin. Here, $\bar{S}(\eta_f)$ is the final degree of polarization of the electron beam. 
Conversely, once $\tau$ is fixed (points $P_4$ and $P_5$ in Figs.~\ref{fig4} (a)-(c)), the MC final results will evolve to a unique $\xi$ value; see Fig.~\ref{fig4}(f).

\begin{figure}[t]
	\begin{center}
		\includegraphics[width=1\linewidth]{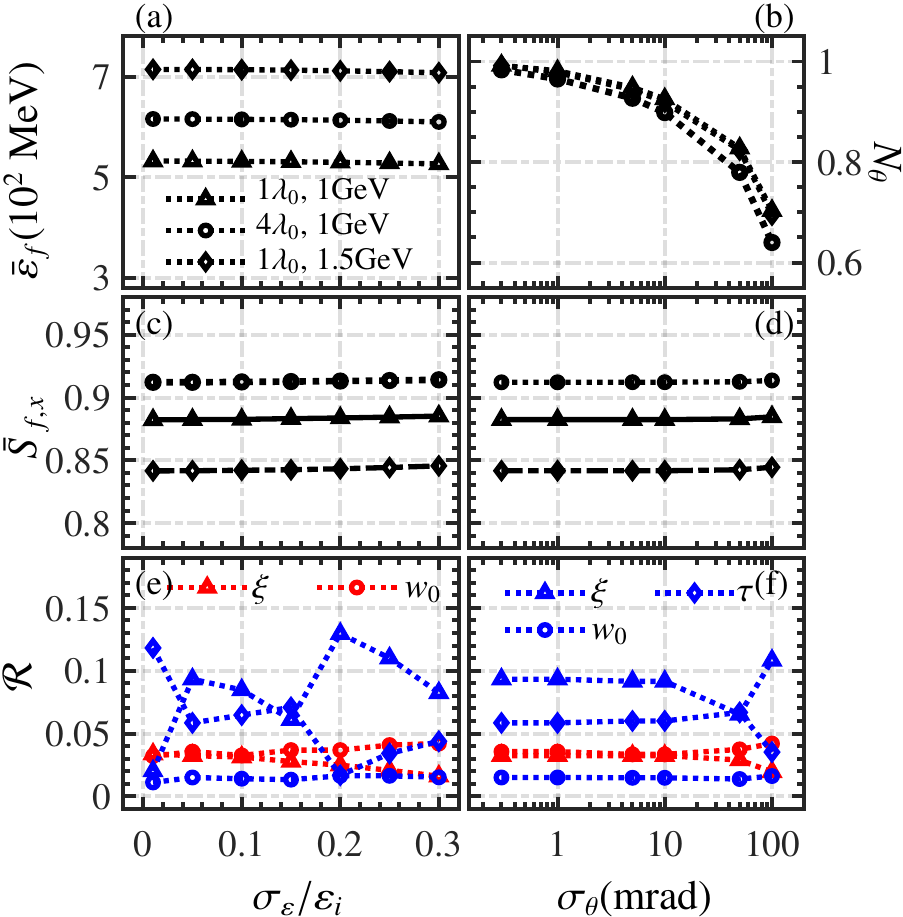}
		\caption{Impact of the probe electron beam parameters on the detection signals. (a) Final average kinetic energies $\bar{\varepsilon}_f$ vs the initial energy spreads $\sigma_\varepsilon/\varepsilon_i$ of the probe electron beams ($\sigma_\theta=0.3$~mrad). Lines marked with triangles, circles, and diamonds denote probe electrons with different beam radii and energies. The initial spin polarization is $\bar{S}_{i,x}=1$ and the laser parameters are the same as in Fig.~\ref{fig4} (d). 
			(b) Relative changes of the angular spread $N_\theta=(\Delta\theta_{f,x}-\Delta\theta_{i,x})/\Delta\theta_{f,x}$ vs the initial angular spread $\sigma_\theta$ ($\sigma_\varepsilon/\varepsilon_i=0.05$) of the probe electron beams, where $\Delta\theta_{i,x}$ and $\Delta\theta_{f,x}$ denote the full-width-at-half-maximum (FWHM) of the initial and final angular spectra along the $x$-direction, and $\theta_x=\arctan(p_x/p_z)$. 
			(c) and (d): Final transverse spin degrees of polarization of the scattered electron beams $\bar{S}_{f, x}$ vs $\sigma_\varepsilon/\varepsilon_i$ and $\sigma_\theta$, respectively. (e) and (f): Relative errors $\mathcal{R}$ vs $\sigma_\varepsilon/\varepsilon_i$ and $\sigma_\theta$, respectively. The red and blue lines are the relative errors from analytical asymptotic estimation and BPNN, respectively. Lines marked with triangles, circles, and diamonds, denote $\mathcal{R}$ of $\xi$, $w_0$, and $\tau$, respectively. }\label{fig5}
	\end{center}
\end{figure}

Compared with the signals from dynamical statistics, the degree of spin polarization is more accurate and more robust with respect to fluctuations in energy and angular spread of the electron beam probe; see Figs.~\ref{fig5} (a)-(d). 
As the initial energy spread $\sigma_\varepsilon/\varepsilon_i$ varies from 1\% to 30\%, the average energy ($\bar{\varepsilon}_f\sim500$ MeV) of the final electron beam ($w_e=1 \lambda_0, \varepsilon_i=1$ GeV) changes by $\sim1\%$; see Fig.~\ref{fig5} (a). 
However, the effect of energy spread on the spin polarization $\bar{S}_{f,x}$ of the final state is about $\sim 0.3\%$; see Fig.~\ref{fig5} (c). 
According to Eq.~(\ref{solution1}) expressing analytical AE, $S_f\sim e^{-k_1\gamma_e}$ and $\delta S_f\sim\delta\gamma_e k_1 e^{-k_1\gamma_e}$, which leads to the conclusion that the spin variations due to dynamics exhibit exponential decay.
Similarly, while the initial angular spread $\sigma_\theta$ changes from 0.3 to 100 mrad, the normalized variation of angular spread $N_\theta$ is $\sim30\%$, and the effect on the spin $\bar{S}_{f,x}$ is $ \sim 0.2\%$. 
In short, detection accuracy of the spin signal is one to two orders of magnitude higher than that of the dynamic signal. 
Relative errors $\mathcal{R}$ of the analytical AE and ML-assisted spin signals are shown in Figs.~\ref{fig5} (e) and (f). Due to angular and energy spread, the relative errors $\mathcal{R}$ of the analytical AE, for $\xi$ and $w_0$, are both kept within 5\%, while the ML-assisted method can simultaneously predict three parameter values for $\xi$, $w_0$ and $\tau$, with relative errors $\mathcal{R}\lesssim10\%$. Especially for $w_0$, accuracy of the ML-assisted method is at least twice as good as that of the analytical prediction.

\section{CONCLUSION}\label{four}
We have put forward an ML-assisted method to diagnose the spatiotemporal properties of an ultrafast ultraintense laser pulse, namely, the pulse duration $\tau$, peak intensity $\xi$ and focal spot size $w_0$, based on the radiative spin-flip effect of the electrons while experiencing strong NCS.
Our trained BPNN can accurately predict the spatiotemporal characteristics of petawatt-level laser systems with relative errors $\lesssim 10\%$. 
The proposed method is accurate and robust with respect to fluctuations in the electron beam parameters, and can be suitably deployed to currently running or planned multi-petawatt-scale laser facilities. 
Accurate measurement of the ultrafast ultraintense laser parameters may pave the way for future strong-field experiments, of importance to laser nuclear physics investigations, laboratory astrophysics studies, and other fields.

\section{ACKNOWLEDGEMENT}
This work is supported by the National Natural Science Foundation of China (Grants Nos. 11874295, 12022506, U2267204, 11905169, 12275209, 11875219, and 12171383), the Open Fund of the State Key Laboratory of High Field Laser Physics (Shanghai Institute of Optics and Fine Mechanics), and the foundation of science and technology on plasma physics laboratory (no. JCKYS2021212008). The work of YIS is supported by an American University of Sharjah Faculty Research Grant (FRG21).

\nocite{*}
\bibliography{library}% Produces the bibliography via BibTeX.

\end{document}